# Electrochemical Surface Modification of a 2D MoS$_2$ Semiconductor

*Martina Lihter[1]\*, Michael Graf[1], Damir Iveković[2], Tzu-Hsien Shen[3], Yanfei Zhao[4], Vasiliki Tileli[3] and Aleksandra Radenovic[1]\**

Martina Lihter, Michael Graf, Aleksandra Radenovic
[1]Laboratory of Nanoscale Biology, Institute of Bioengineering, School of Engineering, EPFL, 1015 Lausanne, Switzerland
E-mail: martina.lihter@epfl.ch, aleksandra.radenovic@epfl.ch

Damir Iveković
[2]Laboratory of General and Inorganic Chemistry and Electroanalysis, Faculty of Food Technology and Biotechnology, University of Zagreb, 10 000 Zagreb, Croatia

Tzu-Hsien Shen, Vasiliki Tileli
[3]Institute of Materials, EPFL, 1015 Lausanne, Switzerland

Yanfei Zhao
[4]Laboratory of Nanoscale Electronics and Structures, Institute of Electrical Engineering, School of Engineering, EPFL, 1015 Lausanne, Switzerland

**Abstract**

The surface modification of 2D semiconducting materials, such as transition metal dichalcogenides (TMDCs), is becoming important for a diverse range of applications, such as biosensing, catalysis, energy generation and energy storage. Due to the chemical inertness of their basal plane, the surface modification of 2D TMDCs is mainly limited to their defective sites, or it requires a conversion of TMDC from its semiconducting into a metallic phase. In this work, we show that the basal plane of a 2D semiconductor molybdenum disulfide (MoS$_2$) can be modified by electrochemical grafting of aryl-diazonium salt, such as 3,5-bis(trifluoromethyl)benzenediazonium tetrafluoroborate. To investigate the applicability of this method, we perform electrografting on MoS$_2$ nanoribbons by addressing them individually via a different electrode. High spatial selectivity of this method on the nanoscale opens the possibility for specific surface modification of neighboring 2D layers and nanostructures that are contacted by electrodes. This method could be potentially applicable to other 2D semiconducting materials that are active in the same potential window in which the electrochemical reduction of aryl diazonium salts occurs.

**Introduction.** Currently, there is a broad interest in 2D transition metal dichalcogenides (TMDCs) for various applications in electronics[1,2], sensing and biosensing[3,4], energy generation[5,6] and energy storage[7]. This interest is motivated by a well-defined two-dimensional crystalline and electronic structure of 2D TMDCs, combined with their exceptional mechanical and chemical stability.[8] However, the chemical inertness of 2D TMDCs becomes an obstacle when it comes to tailoring their surface properties via chemical modifications[9,10]. The basal plane of 2D TMDCs is chemically inert due to the absence of dangling bonds, leaving the edges, grain boundaries, and atom vacancies as the only reactive sites for chemical modifications.[11–13] Therefore, it is rather difficult to obtain the 2D TMDC material whose surface is modified with a uniform and dense molecular layer. Constant progress in the synthesis of the low-defect 2D TMDCs, driven by the need for 2D materials with controlled and reproducible properties, makes the chemical surface modification of 2D TMDCs even more challenging. The surface modification is among the easiest and most effective ways to alter the electronic and optical properties of the low-dimensional nanomaterial and tune the interactions between the material and other chemical species. Therefore, finding the ways to chemically modify the surface of the 2D TMDC materials with homogeneous and uniformly distributed molecular layers is one of the key steps toward practical applications of 2D TMDCs. For many of these applications (e.g. chemical sensing and biosensing, nanoelectronics or nanooptics), tailoring the surface properties of the material on a nanoscale is almost an essential, which is another feature that the surface modification methods should meet in order to be successfully applied in fabrication of the 2D TMDC-based devices.

In this work, we focus on 2D molybdenum disulfide ($MoS_2$), which is a typical representative and one of the most investigated TMDCs. The majority of reported methods on surface modification of 2D $MoS_2$ are wet chemistry methods limited to reactions with defective sites in $MoS_2$, such as sulfur-vacancies[14–20]. Only a few reports showed that it is possible to create a covalent S-C bond on $MoS_2$[21–23] and modify the entire basal plane. However, these methods require either pre-existing defects as nucleation sites[23], or a conversion of semiconducting $MoS_2$ 2H-phase into metallic 1T-phase,[21,22] which is often unpractical and incompatible with the intended application of $MoS_2$. Additionally, non of the above-mentioned methods allow the $MoS_2$ modification to be performed with spatial selectivity, especially on the nanometer scale. An attractive alternative that could enable a uniform basal plane modification, as well as high spatial selectivity, is electrochemically controlled surface modification. $MoS_2$ is an n-type semiconductor, and polarizing it cathodically (toward lower potentials)[24–27] facilitates the electron transfer from its surface to chemical species in solution. This property could be exploited for electrochemical grafting of highly reactive species, such as aryl radicals produced by electrochemical reduction of aryl diazonium salts[28–31], to the entire $MoS_2$ basal plane. The reduction of aryl diazonium ion produces very reactive aryl radicals, which tend to covalently bind to the surface of various materials, including the $MoS_2$[22,23]. Compared to other types of grafting (e.g. spontaneous grafting, grafting controlled by heat, or by the addition of chemical agents etc.), in electrochemical grafting the reaction occurs at the electrode surface, making it highly localized. Electrografting on $MoS_2$ has to be performed at relatively mild potentials for two reasons: first, to avoid possible metallization of $MoS_2$ that might occur at cathodic potentials,[32,33] and second, to avoid hydrogen evolution reaction on $MoS_2$, which limits the useable potential window for $MoS_2$ modification.[34] Aryl diazonium salts have an advantage that they can be reduced at relatively low cathodic potentials (around 0 V vs. standard hydrogen electrode), and as well in aprotic media, which avoids hydrogen evolution reaction. The grafting process is self-limiting, meaning that the grafting terminates after deposition of certain number of molecular layers, and can be easily controllable.[35] In addition, differently substituted aryl diazonium salts can be easily synthesized from corresponding aniline compounds, which enables introducing different chemical groups on the surface of the material. In this work, we perform electrografting of a 3,5-bis(trifluoromethyl)benzenediazonium tetrafluoroborate (TFMB) on high-quality $MoS_2$ nanoribbons controllably grown by metal-organic chemical vapor deposition (MOCVD)[36]. TFMB was chosen as a model aryl diazonium compound in this work since it contains six fluorine atoms per molecule, which greatly facilitates the selective detection of electrografted molecules by electron probe techniques at the nanoscale.

**Results and discussion**

Electrografting of TFMB on 2D $MoS_2$ material was carried out by connecting $MoS_2$ as a working electrode using a three-electrode set-up (**Figure 1a**). To investigate the applicability and selectivity of the electrografting process at the nanoscale, we used devices with electrically connected 2D $MoS_2$ nanoribbons (**Figure 1c**). Nanoribbons of single-layer $MoS_2$, 500 nm in width, each contacted by two metal contacts were designed on a thin $SiN_x$ membrane by nanofabrication process described in detail in **Methods Section**. The device was immersed in a solution of TFMB in acetonitrile with tetrabuthylammonium perchlorate (TBAP) as a supporting electrolyte. Acetonitrile was used as an aprotic solvent to stabilize the radical intermediates of the aryl diazonium salt,[37] as well as to prevent the hydrogen evolution on $MoS_2$. Electrografting was performed by polarizing the working electrode from +450 mV to -750 mV vs. Ag | AgCl. Cathodic polarization initiates the electron transfer from $MoS_2$ surface to TFMB cation in solution, which results in generation of TFMB radical as schematically depicted in **Figure 1b**. The reaction can be described by: $(CF_3)_2C_6H_3N_2^+ + e^- \rightarrow (CF_3)_2C_6H_3 \cdot + N_2$, where $(CF_3)_2C_6H_3N_2^+$ is TFMB cation, and $(CF_3)_2C_6H_3 \cdot$ a corresponding aryl radical. The radical binds to the $MoS_2$ surface forming a covalent S-C bond. **Figure 1c** shows the cyclic voltammogram of the electrografting process. A broad peak in the first scan, around -0.3 V vs. Ag | AgCl, originates from the reduction of the of TFMB cations.[38] The

disappearance of the peak in second and subsequent cycles indicates the formation of a compact organic layer on the electrode surface that blocks further electron transfer between the electrode and the diazonium cations in solution[30,38,39] terminating the process.

To investigate the thickness and the uniformity of the electrochemically deposited film, we performed the atomic force microscopy (AFM) imaging on 2D $MoS_2$ monocrystals. **Figure 2a** shows an optical image of a region with several electrically connected 2D $MoS_2$ monocrystals on which TFMB was electrografted. A part of an electrically connected $MoS_2$ monocrystal before and after the electrografting is shown in **Figure 2b** and **c**, respectively. The line profiles of the corresponding cross-sections (**Figure 2b** and **c**, insets) reveal the local change in height of 2.7 ± 0.4 nm, which corresponds to approximately four to five TFMB molecular layers (the theoretical height of TFMB monolayer is ~6.1Å). **Figure 2d** and **Figure 2e** show a large area (indicated in **Figure 2a**) before and after the electrografting process, respectively. We observe that the height of unconnected monocrystals and the surface roughness has not changed significantly, indicating that the deposition took place only on monocrystals that were connected to the electrode and polarized (more details in **Figure S2** and **Table S1**). In contrast with previously reported chemical grafting on metallic $1T-MoS_2$,[22] which occurs spontaneously on all surfaces exposed, electrochemically controlled grafting is highly localized to the cathodically polarized $MoS_2$ surface where the aryl radicals are produced. The spatial selectivity of the deposition is in this case limited by diffusion path length of the radicals, $l$, which depends on their diffusion coefficient, $D$, and the half-life, $\tau$. The values for $D$ and $\tau$ of aryl radicals are usually on the order of $10^{-6}$ $cm^2$ $s^{-1}$[40] and microseconds[41], respectively, which can be used to calculate $l$ by the relation: $l = \sqrt{D \times \tau}$. The estimated value of diffusion path length is below 100 nm, indicating that the reaction layer around $MoS_2$ is extremely thin, resulting in high spatial selectivity of the electrografting method.

**Figure 2f** shows that the deposited layer thickness increases near the edges and grain boundaries, which can be attributed to two effects. One of them is the hemispheric diffusion at the edges, which increases the flux of TFMB cations at the edges of $MoS_2$. The second one is the higher electron transfer rate occurring on the edges and grain boundaries[27,42]. In 2D $MoS_2$, the edges of the crystal and the sulfur vacancies are sites where molybdenum atoms are in a (+IV) valence state, leaving them coordinately unsaturated. The $d_{xy}$ and $d_{x^2-y^2}$ orbitals of the molybdenum atoms located at the edges extend into the electrolyte and interact strongly with the molecules from the environment (electron-donor species). These sites are thus more reactive, resulting in higher heterogeneous electron transfer rates on the edges, grain boundaries and defects compared to the rest of the basal plane.[24] The activity of the $MoS_2$ basal plane depends as well on the efficiency of injecting the electrons into 2D material, which depends on the quality of the contact between the electrode and the 2D material[43] and the interconnection between the individual monocrystals.[26] It is, therefore, expected that the layer thickness and the local surface roughness will vary as well between the individual $MoS_2$ flakes, just as observed in **Figure 2d**.

For inspecting the selectivity of the method at the nanoscale, we performed the transmission electron microscopy (TEM) imaging on a device with three $MoS_2$ nanoribbons presented in **Figure 3a**. The 20 nm thin $SiN_x$ membrane is transparent to an electron beam enabling the inspection of the deposited layer using TEM. The electrografting was performed on two ribbons by polarizing them cathodically, while the third ribbon was kept as a control and was not polarized. **Figure 3b and Figure 3c** show bright-field TEM (BF-TEM) images of the same device after the electrografting process. Both ribbons subjected to electrografting clearly exhibit an increased mass contrast in comparison to the control ribbon, indicating that the TFMB layer was selectively grafted only on the cathodically polarized nanoribbons. A comparison of the nanoribbons' TEM image contrast before and after the deposition is presented in Figure S3. The difference in image contrast is further qualitatively assessed by plotting the mean intensity profile (**Figure 3c**, line plot) along with a narrow box (**Figure 3c**, blue box), revealing three distinct intensity levels that correspond to the $SiN_x$ membrane, the unmodified $MoS_2$, and the modified $MoS_2$. The observed contrast is relatively uniform along the inspected part of the nanoribbon surface indicating that at the scale of several hundreds of nanometers the entirety of the ribbon is homogeneously modified. It should be noted that the shortest distance between the unmodified and the closest modified ribbon is only about 300 nm, demonstrating high spatial selectivity of this modification method.

To additionally confirm the selectivity of the deposition, we performed an elemental analysis of the modified and the control $MoS_2$ ribbons. We utilized the energy-dispersive X-ray spectroscopy (EDX) in scanning TEM (STEM-EDX) mode. **Figure 3d** shows the high angle annular dark field STEM (HAADF-STEM) image and EDX elemental maps characteristic for the region denoted with a red box in **Figure 3c**. The EDX maps of the modified ribbon show a pronounced carbon and fluorine signal, originating from the $CF_3$-substituted aryl groups grafted on the surface. The EDX spectra can be found in **Supplementary Figure S4b**. The fluorine signal (**Figure 3e**) reveals higher content of fluorine in the modified $MoS_2$ ribbon compared to the control ribbon, which is consistent with previous density functional theory calculations[23] showing that spontaneous grafting on $MoS_2$ is energetically unfavorable. The smaller fluorine peaks, visible in EDX spectra of $SiN_x$ membrane and $MoS_2$ control ribbon, mainly originate from the fluorine chemistry process used in the device fabrication. This was additionally confirmed by comparing the EDX data of the device exposed to TFMB and the reference sample (**Table S2**). The mean value of the F/Si atomic ratio is slightly higher for the $MoS_2$ exposed to the TFMB solution than the reference $MoS_2$, which can be attributed to a non-electrochemical deposition process, such as adsorption.

As previously mentioned, it has been shown that cathodic polarization might induce the phase transition of $MoS_2$ from semiconducting 2H to metallic 1T-phase.[32,33] To check if the metallization of $MoS_2$ occurred during the deposition, we measured the conductance of $MoS_2$ nanoribbons before and after the modification in dry conditions (**Figure S5**). The conductance after the modification got increased for one to two orders of magnitude (**Figure S6**), while the expected conductance increase due to metallization is about seven orders of magnitude.[44] We, therefore, conclude that MoS$_2$ did not get metalized during electrochemical grafting.

**Conclusion**

We have shown that the TFMB aryl-diazonium salt can be electrochemically grafted on the basal plane of 2D semiconducting MoS$_2$. Compared to previously reported MoS$_2$ chemical modification methods, this approach is not limited to defective sites, neither it requires MoS$_2$ phase transition. The electrografting occurs only on MoS$_2$ that is cathodically polarized, making the deposition spatially highly selective at the nanoscale. We demonstrated this by modifying the surface of MoS$_2$ nanoribbons by addressing them individually via a different electrode. This approach paves the way for tailoring surface properties of 2D TMDCs, which could be useful for different applications, particularly in biosensing. A specific functionalization of neighboring ribbons at the nanoscale could increase the sensing capabilities of 2D material-based sensors[45,46]. Sensors based on the ionic and electronic in-plane measurement of nanopores in 2D materials[4] could as well take advantage of this approach by engineering specific interactions between the 2D material and the analyte. Diazonium salts are especially suitable for electrografting since they can be easily reduced, and most importantly, they offer the possibility of introducing a wide variety of different functional groups on materials surface. Surface modification by electrochemical grafting could be potentially applicable as well to other 2D semiconducting materials that can act as an electrode in the same potential window in which the electrochemical reduction of aryl diazonium salts occurs.

**Experimental Section**

All the reagents and solvents used were analytical grade and purchased from Sigma – Aldrich, Merck, Darmstadt, Germany.

*MoS₂ transfer:* 2D MoS$_2$ monocrystals were grown by metal organic chemical vapor deposition on sapphire and characterized as reported previously.[36] The thickness of monocrystals was confirmed by optical contrast and AFM measurements. Monocrystals were transferred onto the target substrate by wet transfer method described elsewhere.[47]

*MoS₂ devices.* After the MoS$_2$ transfer and cleaning, electron beam lithography (EBL) and electron beam assisted metal evaporation was used to pattern and evaporate the electrodes (5 nm Ti / 70 nm Au / 5 nm Pt) to contact MoS$_2$ monocrystals.

*Devices with MoS₂ nanoribbons on SiN$_x$ membrane.* The protocol for the fabrication process is available elsewhere.[4,47] In short, prior to MoS$_2$ transfer, we created 30 µm x 30 µm SiN$_x$ membrane with an aperture (80 nm in diameter) by using EBL, fluorine-based (C$_2$F$_6$) reactive ion etching (RIE), and wet etching in a hot potassium hydroxide solution (30% w/w). After the MoS$_2$ transfer and cleaning, EBL and metal evaporation were used to pattern and evaporate the electrodes (5 nm Ti / 70 nm Au / 5 nm Pt) on top of the MoS$_2$. After patterning these contacts, the MoS$_2$ monolayer is etched into nanoribbons (2 µm x 500 nm) using EBL and RIE (O$_2$ plasma). In all EBL steps, a three-step alignment scheme was used to achieve the necessary precision between individual lithographic steps. Before electrografting, the devices were rinsed with acetone and isopropanol, and carefully dried with a stream of nitrogen. Each device was fixed to a PCB board, the electrodes were wire-bonded to the PCB contacts serving to connect the electrodes to a potentiostat.

*Electrochemical grafting.* Grafting was performed by cyclic voltammetry (CV) in a solution of 3,5-bis(trifluoromethyl)benzenediazonium tetrafluoroborate, TFMB, (5 mM) in acetonitrile with tetrabutylammonium perchlorate, TBAP, (100 mM) which was used as a supporting electrolyte. Details on the synthesis and characterization of TFMB can be found in the Supplemental Information. A three-electrode setup was employed, with a $MoS_2$ nanoribbon working electrode, a Pt-wire counter electrode and a double-junction Ag|AgCl|3M KCl reference electrode (BASinc, West Lafayette, USA) connected to the PalmSens2 potentiostat (Palm Instruments BV, Houten, Netherlands). All electrografting experiments were performed at 5 °C under dimmed light. The $MoS_2$ nanoribbon electrode was polarized between +450 mV and -750 mV at 50 mV s$^{-1}$ during 5 cycles. After deposition, the electrodes and the devices were thoroughly rinsed with acetonitrile, and gently dried with a nitrogen flow.

*Atomic Force Microscopy.* The devices and substrates were imaged using microcantilevers (70 kHz, 2 N m$^{-1}$, Olympus) and an atomic force microscope Asylum Cypher (Oxford Instruments – Asylum Research, Santa Barbara, USA) operating in tapping (AC) mode. The theoretical height of TFMB monolayer covalently bonded to $MoS_2$ was calculated from the molecular structure in Avogadro software and estimated to be 6.1Å. The measured height of the layer will be higher due to the interactions of the molecules with the AFM tip.

*Scanning/Transmission Electron Microscopy.* BF-TEM and HAADF-STEM images of the devices were acquired using a Talos TEM, (ThermoFisher Scientific, Hillsboro, Oregon, USA) operated at an accelerating voltage of 80 kV. EDX elemental maps were recorded in the same microscope in STEM mode by collecting characteristic X-ray signals using two silicon-drift in-column detectors. The EDX data was processed and fitted with Velox software (ThermoFisher Scientific, Hillsboro, Oregon, USA).

*Conductance Measurements in Dry Conditions*: The conductance measurements of contacted $MoS_2$ ribbons were performed by measuring I-V characteristics in a two-terminal configuration in air and at room temperature.

**Acknowledgments**

We thank Prof. Carlotta Guiducci for providing access to a potentiostat, Prof. Andras Kis for the access to AFM and probe-station, and Dr. Ahmet Avsar for useful discussions. This work was financially supported by the Swiss National Science Foundation (SNSF) Consolidator grant (BIONIC BSCGI0_157802) and CCMX project ("Large Area Growth of 2D Materials for device integration").

**Author contributions**

M.L. initiated the idea, performed electrochemical deposition, AFM and TEM imaging. M.L., D.I. and A.R. designed the experiments. M.G. fabricated devices. D.I. proposed the type of diazonium salt used, performed the synthesis and FTIR characterization, and participated in the final reviewing and editing of the manuscript. T.S. performed EDX measurements under V.T supervision. Y.Z. synthesized $MoS_2$ material. A.R. supervised the research. M.L. and M.G. wrote the manuscript with the comments of all the co-authors.

**Conflict of interest**

The authors declare no conflict of interest.

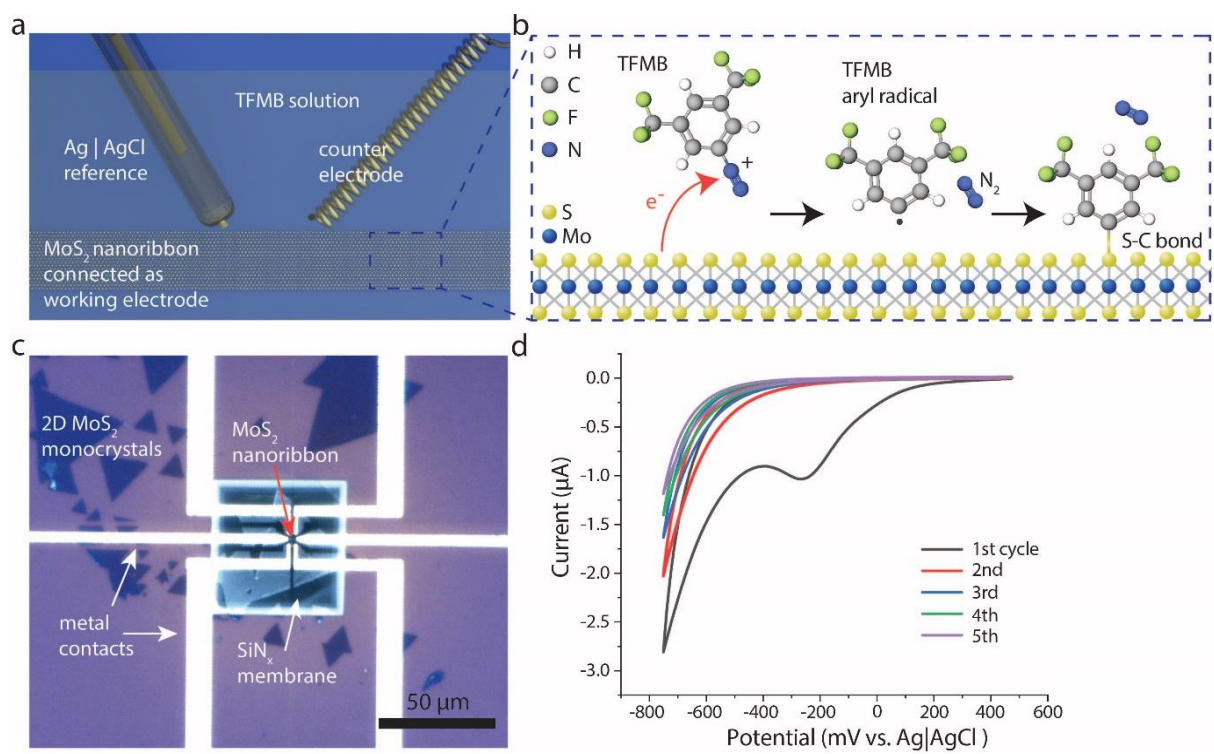

**Figure 1.** (a) An illustration of the experimental set-up for electrochemical grafting. MoS$_2$ is connected as a working electrode, while Ag|AgCl and Pt wire serve as a reference and counter electrode, respectively. The electrodes are immersed in a solution of 3,5-bis(trifluoromethyl)benzenediazonium tetrafluoroborate, TFMB, (5 mM) in acetonitrile with tetrabutylammonium perchlorate, TBAP, (100 mM) which was used as a supporting electrolyte. The objects are not to scale. (b) Schematic of the TFMB electrografting process onto monolayer MoS$_2$. (c) A device with 2D MoS$_2$ monocrystal etched into nanoribbons and fabricated on top of 20 nm thin SiN$_x$ membrane. Each nanoribbon is contacted by one pair of metal contacts. (d) Cyclic voltammogram of TFMB electrografting process. The electrografting was performed in a solution of TFMB (5 mM) and TBAP (100 mM) in acetonitrile. The MoS$_2$ nanoribbon was polarized between +450 mV and -750 mV at 50 mV s$^{-1}$ during 5 cycles.

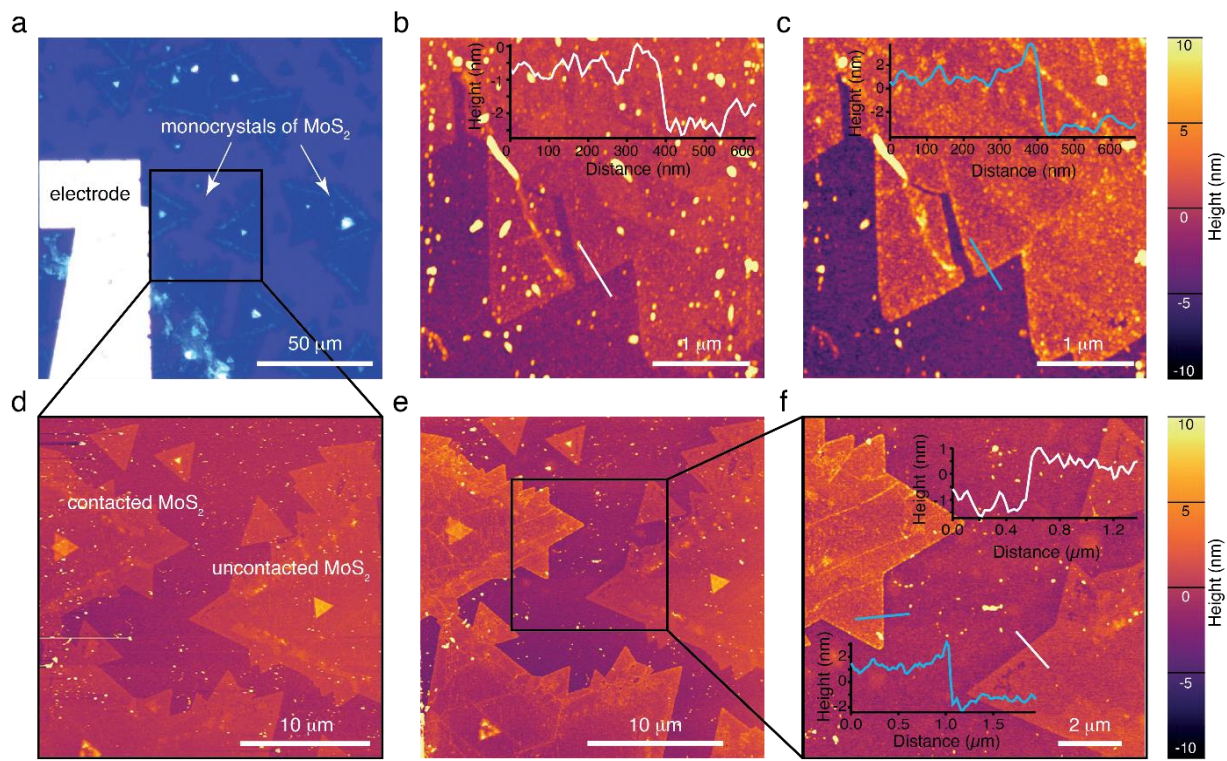

**Figure 2.** (a) Optical image of MoS$_2$ monocrystals contacted by an electrode. The AFM image of a representative MoS$_2$ region before (b) and after (c) electrografting. The insets show the line profile of the corresponding cross-sections before and after the deposition. (d) The AFM image of the area indicated in (a) prior to electrografting, displaying mostly monolayer MoS$_2$. (e) The AFM image of the same area as in (d) after the electrografting process. (f) A close-up view of the region indicated in (e). The insets show the line profile of the two cross-sections represented by a light-blue (modified MoS$_2$) and a white (unmodified MoS$_2$) line.

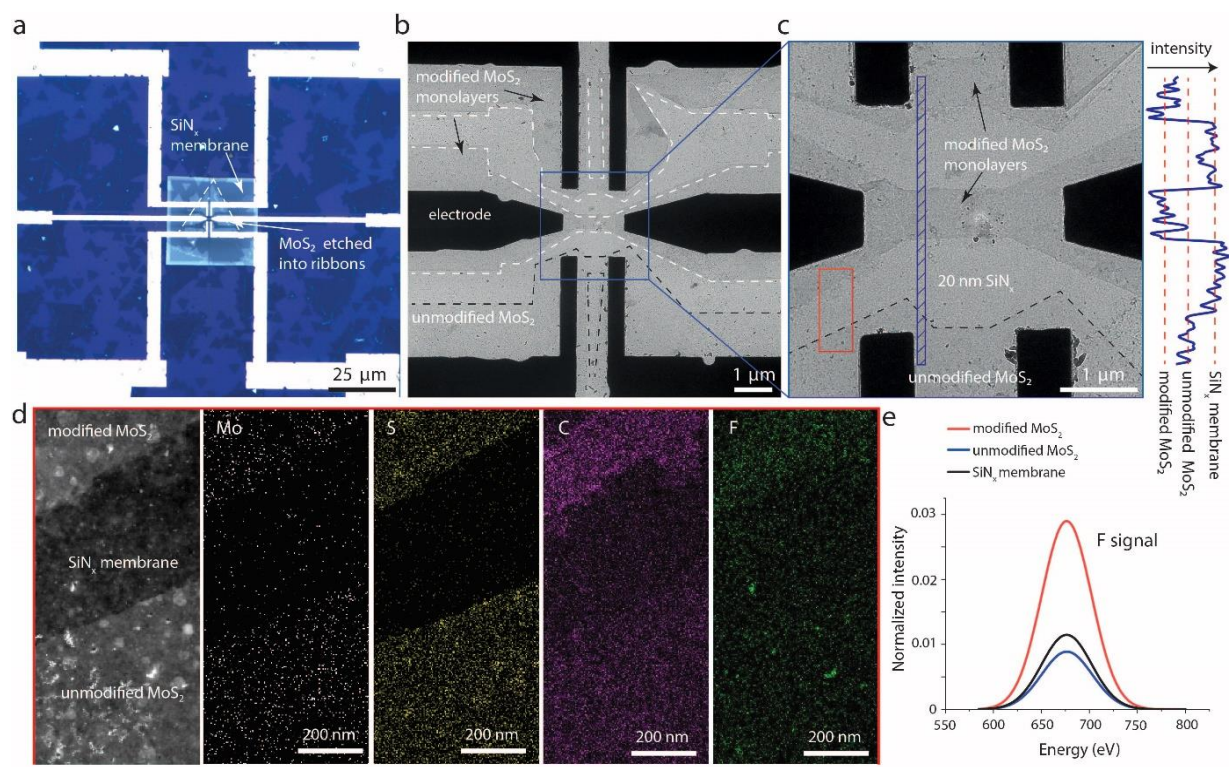

**Figure 3.** (a) The optical image of a device used with $MoS_2$ monolayer transferred onto the $SiN_x$ membrane. The monolayer is etched into three ribbons and contacted by the electrodes. (b) The BF-TEM image of $MoS_2$ ribbons. The upper and middle ribbon have been electrografted with TFMB (5 mM) in acetonitrile with TBAP (100 mM) as a supporting electrolyte. The electrografting was performed during 5 cycles from +450 mV to -750 mV at 50 mVs$^{-1}$. (c) The close-up view of the region in (b) indicated by the blue square. The line profile of the blue area is depicted next to the BF-TEM image. (d) The HAADF-STEM image and EDX elemental maps of Mo, S, C and F, respectively, of the region indicated by a red square in (c). The maps show the net (background corrected and fit) intensities of the elements present. (e) Fluorine EDX spectra obtained from three neighboring regions indicated in (d). F signal detected at the $SiN_x$ membrane and unmodified $MoS_2$ surface mainly originates from the step in the device fabrication process based on fluorine chemistry. The spectra were normalized with respect to the Si signal. The entire EDX spectra can be found in Figure S3 in the SI.

# Supporting Information

**Electrochemical Surface Modification of a 2D MoS$_2$ semiconductor**

*Martina Lihter[1]\*, Michael Graf[1], Damir Iveković[2], Tzu-Hsien Shen[3], Yanfei Zhao[4], Vasiliki Tileli[3] and Aleksandra Radenovic[1]\**

*Aryl diazonium salt synthesis.* 3,5-bis(trifluoromethyl)benzenediazonium tetrafluoroborate (TFMB) was prepared according to the following procedure: 200 µL (1.5 mmol) of 3,5-bis(trifluoromethyl)aniline (TFMA) (98%, Sigma Aldrich, Merck, Darmstadt, Germany) was dissolved in a mixture of 1 mL of tetrafluoroboric acid (48% in water, Sigma Aldrich, Merck, Darmstadt, Germany) and 2 mL of water. After cooling on ice, a solution of 105 mg of sodium nitrite (Sigma Aldrich, Merck, Darmstadt, Germany) in 200 µL of water was added drop by drop to the TFMA solution under stirring. After precipitation, the diazonium salt was vacuum filtered, and rinsed with ice cold water.

*Infrared spectroscopy.* The FTIR spectra were recorded on a PerkinElmer Spectrum Two FTIR spectrometer operated in the attenuated total reflectance (ATR) mode (diamond ATR prism).

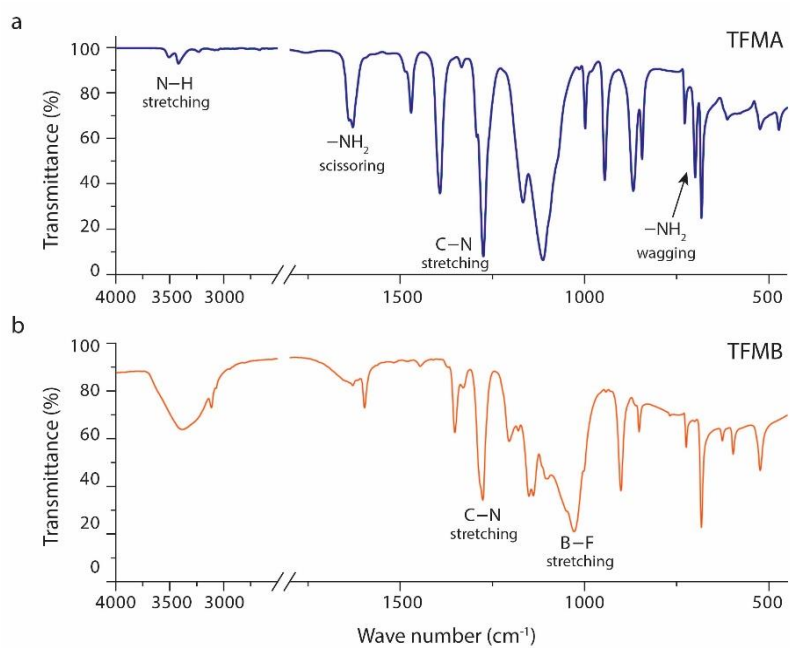

**Figure S1.** (a) FTIR spectrum of 3,5-bis(trifluoromethyl)aniline (TFMA). The bands at 1630 cm$^{-1}$ (-NH$_2$ scissoring), 700 cm$^{-1}$ (-NH$_2$ wagging) and 3300-3500 cm$^{-1}$ (-NH$_2$ stretching) are characteristic for primary amino-group. (b) FTIR spectrum of synthesized 3,5-bis(trifluoromethyl)benzenediazonium tetrafluoroborate (TFMB). Vibration bands characteristic to the -NH$_2$ group do not appear in the spectra of TFMB since the amino-group of TFMA is converted to the diazo-group. A strong band appearing at 1030 cm$^{-1}$ originates from B-F stretching in BF$_4^-$, indicating the formation of diazonium salt[48,49].

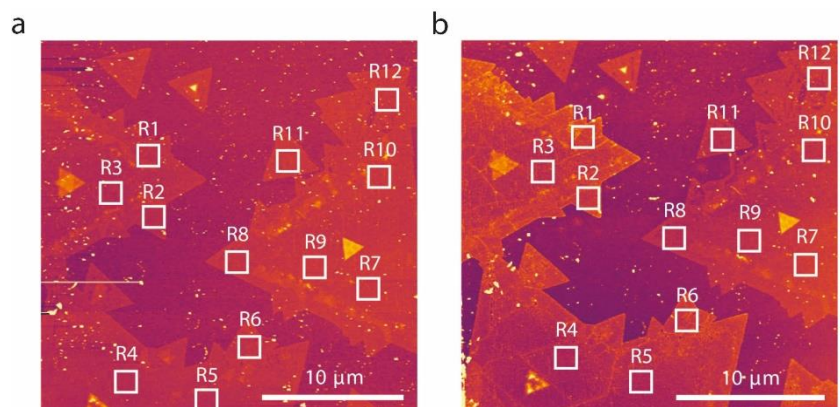

**Figure S2.** AFM micrographs of MoS$_2$ monocrystals recorded before (a) and after (b) of the electrografting of TFMB. Regions (1.5µm x 1.5µm) in which the roughness was inspected are indicated with white squares. Regions R1-R6 are situated on the MoS$_2$ flakes contacting the electrode (directly or through other MoS$_2$ flakes). Neighboring regions R7-R12 have no contact with the electrode. The roughness values are given in Table S1.

**Table S1.** Surface roughness ($R_q$) values of the regions indicated in Figure S2, measured before and after the electrografting of TFMB.

| Region | BEFORE $R_q$ / nm | AFTER $R_q$ / nm |
|---|---|---|
| R1 | 0.54 | 1.21 |
| R2 | 0.32 | 0.84 |
| R3 | 0.50 | 0.80 |
| R4 | 0.29 | 0.57 |
| R5 | 0.32 | 0.59 |
| R6 | 0.45 | 1.04 |
| $R_{mean}$ ± STD | 0.40±0.09 | 0.84±0.23 |
| R7 | 0.35 | 0.37 |
| R8 | 0.48 | 0.49 |
| R9 | 0.38 | 0.36 |
| R10 | 0.46 | 0.45 |
| R11 | 0.75 | 0.68 |
| R12 | 0.77 | 0.74 |
| $R_{mean}$ ± STD | 0.53±0.17 | 0.52±0.15 |

R1-R6 are regions on the MoS$_2$ flakes that were in contact with the electrode, directly or through other MoS$_2$ flakes. Regions R7-R12 had no contact with the electrode. Surface roughness ($R_q$) is expressed as a root mean square deviation of the assessed profile. The mean value of roughness and the corresponding standard deviation for each data set is given in the row labelled as "$R_{mean}$ ± STD". The data sets from regions (R7 - R12) were tested for the hypothesis that the roughness ($R_q$) of the MoS$_2$ flakes that were in no contact with the electrode changed after the electrografting process, i.e. that the grafting of TFMB occurred as well on the unconnected MoS$_2$ flakes. From the *t*-test (paired, two-tailed distribution) a value of p=0.259, was obtained, indicating that the means of the two data sets do not differ significantly. This strongly indicates that the grafting of TFMB occurred only on the MoS$_2$ flakes in contact with the electrode.

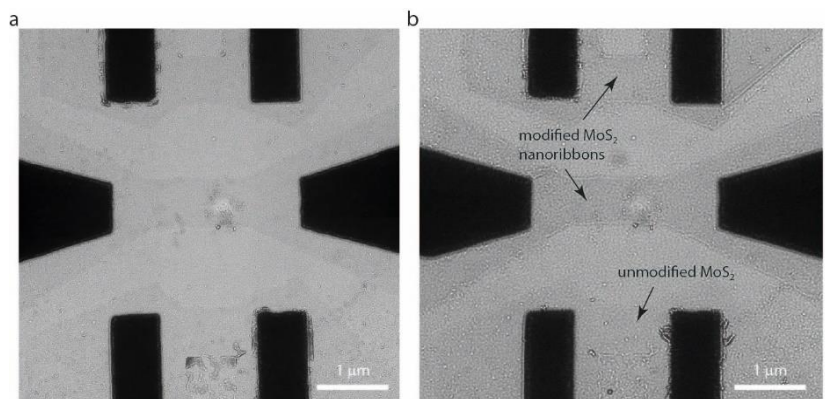

**Figure S3.** TEM images of the MoS$_2$ nanoribbons before (a) and after electrografting modification of two upper ribbons (b). The images are out of focus to emphasize the contrast of MoS$_2$ that in this condition appears slightly darker than the supporting SiN$_x$ membrane. The imaging was performed at an accelerating voltage of 80 kV.

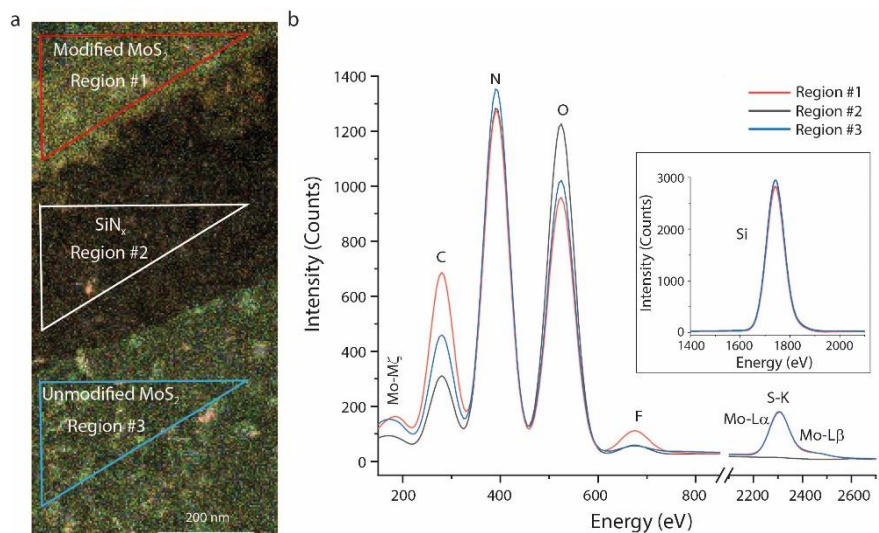

**Figure S4.** (a) The HAADF-STEM image of three regions of interest from which the EDX spectra were taken. (b) Fitted EDX spectra of the regions indicated in (a). The modified MoS$_2$ nanoribbon (Region #1) exhibits an increased content of F and C originating from the electrografted TFMB molecules. F signal of the modified MoS$_2$ surface is significantly higher than the signal from the unmodified MoS$_2$ (Region #3) and the SiN$_x$ surface (Region #2). The higher level of O in Region #2 compared to Region #1 and Region #3, originates most probably from the process of reactive ion etching (RIE) of MoS$_2$ into ribbons by oxygen. The inset shows the level of Si, which is approximately the same on all surfaces.

**Table S2**. The atomic composition for F and Si obtained from EDX data on as-fabricated (reference) device, and on a device that was exposed to TFMB solution.

| REFERENCE |
|---|

|  | MoS$_2$ | | |  | SiN$_x$ membrane | | |
|---|---|---|---|---|---|---|---|
|  | F at% | Si at% | Atomic ratio F / Si |  | F at% | Si at% | Atomic ratio F / Si |
| Region#1 | 0.63 | 37.99 | 0.0166 | Region#9 | 0.21 | 27.95 | 0.0075 |
| Region#2 | 0.13 | 31.03 | 0.0042 | Region#10 | 0.63 | 39.36 | 0.0160 |
| Region#3 | 0.08 | 29.5 | 0.0027 | Region#11 | 0.40 | 38.44 | 0.0104 |
| Region#4 | 0.09 | 31.77 | 0.0028 | Region#12 | 0.26 | 31.79 | 0.0082 |
| Region#5 | 0.35 | 40.5 | 0.0086 | Region#13 | 0.60 | 38.95 | 0.0154 |
| Region#6 | 0.79 | 38.6 | 0.0205 | Region#14 | 0.13 | 33.01 | 0.0039 |
| Region#7 | 0.07 | 30.36 | 0.0023 | Region#15 | 0.26 | 31.66 | 0.0082 |
| Region#8 | 0.10 | 31.52 | 0.0032 | Region#16 | 0.32 | 28.58 | 0.0112 |
| $A_{mean} \pm$ STD |  |  | 0.0076±0.0071 | $A_{mean} \pm$ STD |  |  | 0.0101±0.0041 |
| DEVICE EXPOSED TO TFMB | | | | | | | |
|  | MoS$_2$ | | |  | SiN$_x$ membrane | | |
| Region#17 | 0.37 | 38 | 0.0097 | Region#25 | 0.59 | 38.2 | 0.0154 |
| Region#18 | 0.52 | 37.36 | 0.0139 | Region#26 | 0.52 | 40.13 | 0.0129 |
| Region#19 | 0.49 | 40.23 | 0.0122 | Region#27 | 0.67 | 38.96 | 0.0172 |
| Region#20 | 0.57 | 38.19 | 0.0149 | Region#28 | 0.47 | 38.32 | 0.0123 |
| Region#21 | 0.48 | 39.64 | 0.0121 | Region#29 | 0.47 | 40.19 | 0.0117 |
| Region#22 | 0.57 | 38.84 | 0.0147 | Region#30 | 0.49 | 39.89 | 0.0123 |
| Region#23 | 0.39 | 42.26 | 0.0092 | Region#31 | 0.52 | 40.53 | 0.0128 |
| Region#24 | 0.48 | 42.6 | 0.0113 | Region#32 | 0.56 | 38.66 | 0.0145 |
| $A_{mean} \pm$ STD |  |  | 0.0122±0.0022 | $A_{mean} \pm$ STD |  |  | 0.0136±0.0019 |

Measurements were done at different regions with MoS$_2$ (Region#1-#8 and Region#17-#24) and SiN$_x$ (Region#9-#16 and Region #25-#32). The mean value of the atomic ratio and the corresponding standard deviation for each data set is given in the row labeled as "$A_{mean} \pm$ STD". The presence of fluorine on the reference sample originates from the step in the fabrication process (fluorine chemistry RIE etching). Since this step was done prior to MoS$_2$ transfer, SiN$_x$ regions with no MoS$_2$ exhibit higher content of F. Both, MoS$_2$ and SiN$_x$, exhibit slightly increased F/Si ratio after the exposure to TFMB.

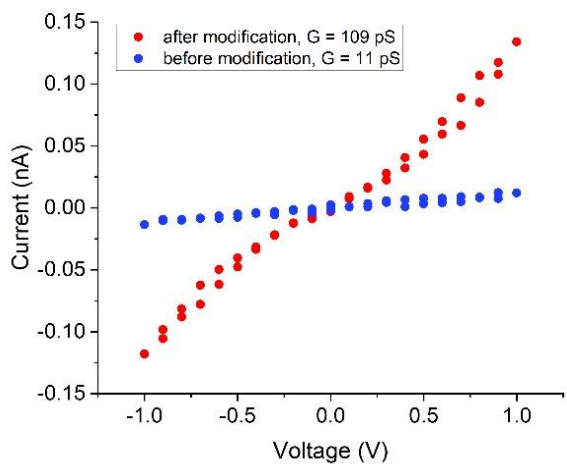

**Figure S5.** I-V characteristics of the same MoS$_2$ nanoribbon measured in air before and after the electrografting with TFMB. The non-linearity of the I-V curve observed after modification is characteristic for the Schottky barrier at the metal-semiconductor junction.[50] The conductance before the modification is consistent with previously reported values by Graf et al.[4] obtained with MoS$_2$ nanoribbons of the same dimensions, and fabricated with the same process flow. In vacuum, these kinds of ribbons can typically exhibit conductance of up to 2 µS.[4] All measurements were done in a two-terminal configuration at room temperature.

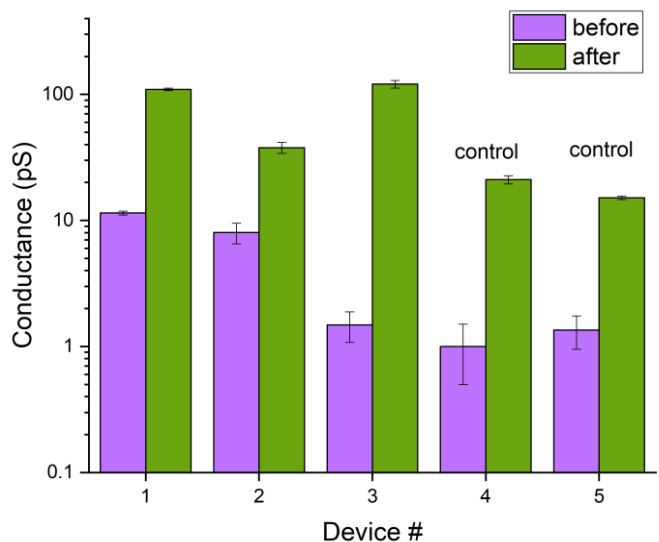

**Figure S6.** The conductance of the devices, measured in air before and after the electrografting of TFMB. The conductance values before the modification are consistent with previously reported values by Graf et al.[4] obtained with MoS$_2$ nanoribbons of the same dimensions, and fabricated with the same process flow. Control devices were not subjected to electrografting, but they were exposed to the solution of TFMB. The enhanced conductance after the exposure to the TFMB solution can be attributed to the reversible physisorption of aromatic molecules on MoS$_2$.[51] All measurements were done in a two-terminal configuration, at room temperature and in air.